\documentclass{nature}
\usepackage{psfig}
\def\simlt{\mathrel{\hbox{\rlap{\hbox{\lower4pt\hbox{$\sim$}}}\hbox{$<$}}}}
\def\simgt{\mathrel{\hbox{\rlap{\hbox{\lower4pt\hbox{$\sim$}}}\hbox{$>$}}}}

\def\M{{\rm M}_{0.3}}
\def\msun{{\rm M}_\odot}
\def\M{{\rm M}_{0.3}}

\title{A common mass scale for satellite galaxies of the Milky Way}

\author{Louis E. Strigari$^{1}$, James S. Bullock$^{1}$, Manoj
Kaplinghat$^{1}$, Joshua~D.~Simon$^{2}$, Marla~Geha$^{3}$, 
Beth~Willman$^{4}$, Matthew~G.~Walker$^{5}$}

\begin{document}
\spacing{1}
\maketitle
\begin{affiliations}
\item Center for Cosmology, Department of Physics and Astronomy, 
University of California, Irvine, CA 92697-4574, USA 
\item Department of Astronomy, California Institute of Technology, 1200 E.
California Blvd., MS105-24, Pasadena, CA 91125, USA
\item Department of Astronomy, Yale University, P.O. Box 208101, New Haven, CT
06520-8101, USA
\item Harvard-Smithsonian Center for Astrophysics, 60 Garden St. Cambridge, MA
02138, USA
\item Institute of Astronomy, University of Cambridge, Madingley Road,
Cambridge CB3 0HA
\end{affiliations}

\begin{abstract}
The Milky Way has at least twenty-three known satellite galaxies that
shine with luminosities ranging from about a thousand to a billion
times that of the Sun. Half of these galaxies were discovered
~\cite{Willman2005,Belokurov2007} in the past few years in the Sloan
Digital Sky Survey, and they are among the least luminous
galaxies in the known Universe. A determination of the mass of these
galaxies provides a test of galaxy formation at the smallest
scales~\cite{Mateo1998,Gilmore2007} and probes the nature of the dark
matter that dominates the mass density of the Universe
~\cite{Spergel2007}. Here we use new measurements of the velocities of
the stars in these galaxies~\cite{SimonandGeha,Walker2007} to show that
they are consistent with them having a common mass of about $10^7$
M$_\odot$ within their central 300 parsecs. This result demonstrates
that the faintest of the Milky Way satellites are the most dark
matter-dominated galaxies known, and could be a hint of a new scale
in galaxy formation or a characteristic scale for the clustering of
dark matter.
\end{abstract}

Many independent lines of evidence strongly argue for the presence of
dark matter in galaxies, clusters of galaxies, and throughout the
observable Universe ~\cite{Spergel2007}. Its identity, however,
remains a mystery. The gravity of dark matter overwhelms that of the
normal atoms and molecules and hence governs the formation and
evolution of galaxies and large-scale
structure~\cite{Peebles1982,White1983,Blumenthal1984}.  In the
currently favored dark matter models, structure in the Universe forms
hierarchically with smaller gravitationally bound clumps of dark
matter -- haloes-- merging to form progressively larger objects.

The mass of the smallest dark matter halo is determined by the
particle properties of dark matter. Dark matter candidates
characterized as cold dark matter can form haloes that are many orders
of magnitude smaller than the least luminous haloes that we infer from
observations.  Cosmological simulations of cold dark matter predict
that galaxies like the Milky Way should be teeming with thousands of
dark matter haloes with masses $\sim 10^6 \msun$, with a steadily
increasing number as we go to the smallest masses
~\cite{Klypin1999,Moore1999,Diemand2005,Diemand:2006ik}.  A large
class of dark matter candidates characterized as ``warm" would predict
fewer of these small haloes ~\cite{Bode:2000gq}.  However, even for
cold dark matter it is uncertain what fraction of the small dark
matter haloes should host visible galaxies, as the ability of gas to
cool and form stars in small dark matter haloes depends on a variety
of poorly-understood physical processes
~\cite{Efstathiou1992,Kauffmann1993,Bullock2001,Kravtsov2004,Mayer2007}.

The smallest known galaxies hosted by their own dark matter haloes are
the dwarf spheroidal satellites of the Milky Way
~\cite{Mateo1998,Gilmore2007}.  These objects have very little gas and
no signs of recent star formation. The least luminous galaxies were
recently discovered in the Sloan Digital Sky Survey
(SDSS)~\cite{Willman2005,Belokurov2007} and follow-up observations
have revealed them to be strongly dark matter
dominated~\cite{Munoz2006,SimonandGeha,Martin2007}.

We have compiled line-of-sight velocity measurements of individual
stars in 18 of the 23 known dwarf galaxies in the Milky
Way~\cite{Walker2007,SimonandGeha}.  We use these measurements to
determine the dynamical mass of their dark matter haloes using a
maximum likelihood analysis~\cite{StrigariRedefining}.  The dynamical
mass is best constrained within the stellar extent, which corresponds
to an average radius of approximately 0.3 kiloparsecs (kpc) for all
the satellites.  We determine this mass, $\M$, by marginalizing over a
five-parameter density profile for dark matter that allows for both
steep density cusps and flat cores in the central regions. It is
important to note that the observed velocity dispersion of stars is
determined by both the dynamical mass and the average anisotropy of
the velocity dispersion (that is, difference between tangential and
radial dispersion).  The anisotropy is unknown and hence we
marginalize over a three-parameter stellar velocity anisotropy
function that allows us to explore a range of orbital models for the
stars~\cite{StrigariRedefining}.

\begin{figure}
%\centering
\centerline{\psfig{figure=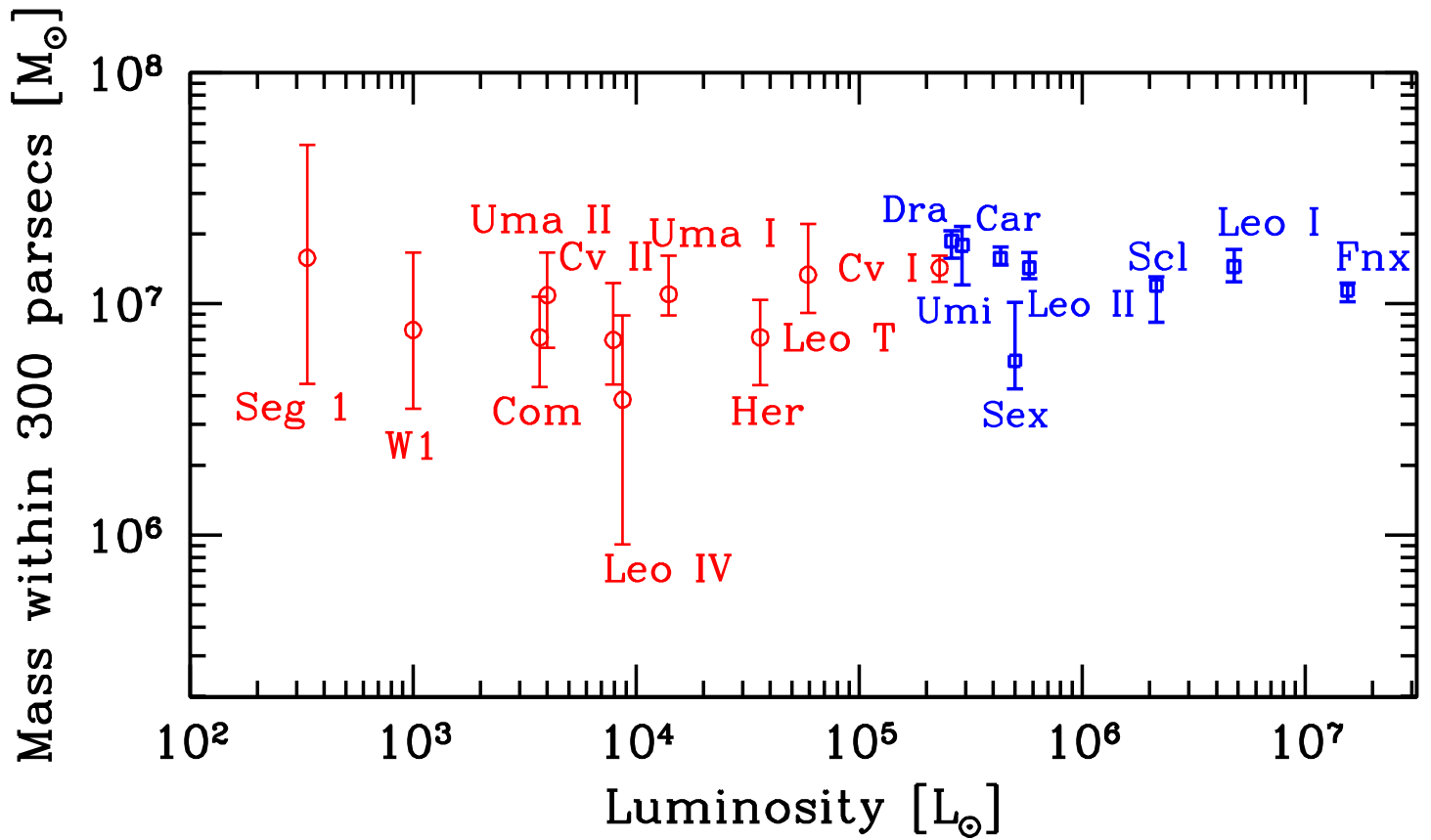}}
\caption{The integrated mass of the Milky Way dwarf satellites, in
units of solar masses, within their inner 0.3 kpc as a function of
their total luminosity, in units of solar luminosities.  The circle
(red) points on the left refer to the newly-discovered SDSS
satellites, while the square (blue) points refer to the classical
dwarf satellites discovered pre-SDSS.  The error bars reflect the
points where the likelihood function falls off to 60.6\% of its peak
value.
\label{fig:ML}
}
\end{figure}

Figure~\ref{fig:ML} shows the resulting determination of $\M$. We find
that all 18 dwarf galaxies are consistent with having a dynamical mass
of $10^7$ solar masses within 0.3 kpc of their centre, despite the
fact that they have luminosity differences over four orders of
magnitude. This result implies a dark matter central density of $\sim
0.1 \, \msun/{\rm pc}^3$ in these galaxies.  Earlier studies suggested
that the highest-luminosity dwarf galaxies all shared a common
mass~\cite{Mateo1993,Gilmore2007}.  With larger stellar data sets,
more than double the number of dwarf galaxies, and more detailed mass
modeling, our results confirm this earlier suggestion and conclusively
establish that the dwarf galaxies of the Milky Way share a common mass
scale.

Because of the proximity of the dwarf galaxies to the Milky Way, it is
possible that tidal effects could change the velocities of stars and
thus affect the mass measurements. In the kinematic data, tidal forces
could be revealed as a velocity gradient across the observed plane of
the dwarf~\cite{Piatek95,Fellhauer:2006jr}.  We have tested the dwarf
galaxies for velocity gradients and have found no conclusive evidence
of tidal effects (see Supplementary Information).

We fitted a $\M$-luminosity relation to the data and obtained $\M
\propto {\rm L}^{0.03 \pm 0.03}$.  This result does not change
significantly if we use the luminosity contained within 0.3 kpc rather
than the total luminosity.  The common mass scale of $\sim 10^7 \msun$
may thus reflect either a plummeting efficiency for galaxy formation
at this mass scale, or that dark matter haloes with lower masses
simply do not exist.

The characteristic density of $0.1 \, \msun/{\rm pc}^3$ may be
associated with a characteristic halo formation time.  In theories of
hierarchical structure formation, the central density of dark matter
haloes is proportional to the mean density of matter in the Universe
when the halo formed. The earlier the formation, the higher the
density.  For cold dark matter models, our measurement implies that
these haloes collapsed at a redshift greater than about 12, or earlier
than 100 million years after the Big Bang.  Measurements of the cosmic
microwave background~\cite{Dunkley:2008ie} suggest that the Universe
went from being neutral to ionized at redshift $11 \pm\ 1.4$. These
dark matter haloes thus formed at approximately the same time that the
Universe was re-ionized.

Within the context of the cold dark matter theory, high-resolution
cosmological simulations can be used to relate the mass within the
central regions of the dark matter halo to the depth of the
gravitational potential well~\cite{BullockCvir}. Simulations show that
$\M \approx 10^{7} \msun \, ({\rm M}_{\rm total}/10^9 \msun)^{0.35}$,
where ${\rm M}_{\rm total}$ is the mass of the halo before it was
accreted into the Milky Way host potential.  Thus, it is possible that
the implied total mass scale of $10^9 \msun$ reflects the
characteristic scale at which supernova feedback~\cite{DekelSilk} or
the imprint of the re-ionization of the
universe~\cite{Bullock2001,Wyithe2006} could sharply suppress star
formation.
 
Perhaps a more speculative, but certainly no less compelling,
explanation of the common mass scale is that dark matter haloes do not
exist with M$_{0.3}$ below $\sim 10^7$ M$_\odot$. This implies that
these dwarf galaxies inhabit the smallest dark matter haloes in the
Universe. Warm dark matter particles have larger free streaming length
than standard cold dark matter particles, which implies that density
perturbations are erased below a characteristic length scale,
resulting in a higher minimum mass for dark matter haloes.  A thermal
warm dark matter candidate with mass of approximately $1 \, {\rm keV}$
would imply a minimum halo mass of $10^9 \msun$. Thus, our mass
determinations rule out thermal warm dark matter candidates with
masses less than about $1 \, {\rm keV}$, but dark matter masses
somewhat larger than 1 keV would yield a minimum dark matter halo mass
consistent with the mass scale we observe.

Future imaging surveys of stars in the Milky Way will provide a more
complete census of low-luminosity Milky Way satellites, with the
prospects of determining whether astrophysics or fundamental dark
matter physics is responsible for setting the common mass scale.  In
particular, the masses for the faintest dwarf galaxies will become
more strongly constrained with more line-of-sight velocity data. This
will sharpen the observational picture of galaxy formation on these
small-scales and provide data around which theories of galaxy
formation may be built.

\begin{addendum}
\item We thank Kathryn Johnston and Simon White for discussion on this
paper.  We thank Jay Strader for help in the acquisition of data for
Segue 1 and Willman 1.
\item[Competing Interests] The authors declare that they have no
competing financial interests.  
\item[Correspondence] Correspondence 
should be addressed to Louis Strigari~(email: lstrigar@uci.edu).
\end{addendum}

\clearpage

\clearpage

\def\simlt{\mathrel{\hbox{\rlap{\hbox{\lower4pt\hbox{$\sim$}}}\hbox{$<$}}}}
\def\simgt{\mathrel{\hbox{\rlap{\hbox{\lower4pt\hbox{$\sim$}}}\hbox{$>$}}}}
\def\si{\sigma_{m,i}} \def\st{\sigma_{t,i}} \def\sB{\sigma_{B}}
\def\stB{\sigma_{tB}}
%%% end local macros     

%\bibliographystyle{naturemag}

%\title{Supplementary Information}

%\begin{document}

%\maketitle

{\bf {\Large Supplementary Information}} 

In this section we present the details of our dynamical mass modeling. 
We begin by reviewing the Jeans formalism for determining the mass 
of a galaxy using line-of-sight velocities. We then discuss our specific treatment of the 
line-of-sight velocity  data and introduce the likelihood function that we use in our analysis. 
Next, we present our results and discuss the implications. Finally, we discuss
the systematics that may affect the mass modeling. 

%%% Methodology section
\section{Mass modeling}
\label{sec:mmassmodeling}
When modeling the stellar distribution of a dwarf galaxy in close
proximity to the Milky Way (MW), the internal gravitational force from
the dwarf must be compared to the external force from the Milky Way.
Generally, the internal gravitational force from the dwarf is  $\sim
\sigma^2/R_s$, where $\sigma$ is the velocity dispersion and $R_s$ is
the half-light radius of the dwarf.   The external tidal force from the MW
potential is $\sim (220 \, {\rm km}/{\rm s})^2R_s/D^2$,  where $D$ is
the distance to the dwarf from the center of the MW, and $220 \, {\rm
km}/{\rm s}$ is the approximate rotation speed of the MW in the outer
regions of the halo where the dwarfs are located. The most luminous of
the MW dwarfs have half-light radii of $R_s \sim 400$ pc, and the least
luminous of the dwarfs have half-light radii of $R_s \sim 10-100$ pc. The
velocity dispersions vary in the range $\sigma \sim 5-15$  km
s$^{-1}$.  Comparing the internal and external forces on dwarfs in the
observed distance range of $\sim 20-250$ kpc, we find that the
internal gravitational  forces are typically larger by $\sim
100$. Note that this estimate does not exclude the possibility that 
the dwarfs have been  tidally stripped in the past; 
it does, however, allow us to proceed safely with the
assumption that the surviving stellar distributions trace the local potential.

Previous studies of some Milky Way dwarf spheroidals (dSphs)  show
that these objects exhibit no streaming motions
~\cite{Walker2006,Koch2007}.  Making the further assumption that these
systems are in  steady state, the three-dimensional radial  velocity
dispersion of the system, $\sigma_{r}$, is given  by the Jeans
equation
\begin{equation} 
r \frac{d(\rho_{\star} \sigma_r^2)}{dr}  =  - \rho_{\star}(r) \frac{G M(r)}{r}
- 2 \beta(r) \rho_{\star} \sigma_r^2.
\label{eq:jeans}         
\end{equation}
This equation is valid if the potential of the system is
spherically-symmetric, which has been shown to be a good  description
of dark matter satellites in a host potential~\cite{Kuhlen2007}. To
convert to the  observable quantity, we must integrate the solution to
the  Jeans equation along the line-of-sight. Performing this
integration gives~\cite{BinneyTremaine}
\begin{equation} 
\sigma_t^{2}(R) = \frac{2}{I_\star(R)} \int_{R}^{\infty}  \left ( 1 -
\beta \frac{R^{2}}{r^2} \right ) \frac{\rho_{\star} \sigma_{r}^{2}
r}{\sqrt{r^2-R^2}} dr.
\label{eq:sigmaLOS}
\end{equation} 
Here $R$ is defined as the projected radius in the plane  of the
sky. The three-dimensional stellar
density profile is defined by $\rho_\star(r)$, which is  determined
from the projected stellar distribution, $I_\star(R)$.  The stellar
velocity anisotropy is defined as  $\beta = 1 -
\sigma_\theta^2(r)/\sigma_r^2(r)$,  where $\sigma_\theta$ is the
tangential component of the  velocity dispersion. We now discuss in
turn our  parameterizations of the different functions entering in
equation~\ref{eq:sigmaLOS}. 

\subsection{Stellar Surface Density}
It is standard to fit the stellar surface densities of the  systems we
study to either Plummer or King profiles.  The surface density for the
King profile is~\cite{King1962}
\begin{equation}
I_{{\rm king}}(R)  = k\left [ \left ( 1 + \frac{R^2}{r_c^2} \right
)^{-1/2} -   \left ( 1 + \frac{r_{lim}^2}{r_c^2} \right )^{-1/2}
\right ]^2,
\label{eq:king}
\end{equation}
which results in a de-projected three-dimensional density of
\begin{eqnarray}
\rho_{{\rm king}}(r) = \frac{k}{\pi r_c [1+(r_{lim}/r_c)^2]^{3/2}z^2}
\left [ \frac{1}{z} \cos^{-1} z - \sqrt{1-z^2} \right ],
\label{eq:rhoking}
\end{eqnarray}
where $z^2 = (1+r^2/r_c^2)/(1+r_{lim}^2/r_c^2)$.  The normalization
constant, $k$, for the King profile thus is  irrelevant when applying
the Jeans equations.  The King profile depends on two parameters,
$r_{lim}$ and $r_c$.

The surface density for the Plummer profile is given by
\begin{equation}
I_{{\rm pl}}(R) =  \frac{4}{3} \frac{\rho_0 r_{\rm
pl}}{\left[1+(R/r_{\rm pl})^2\right]^2},
\end{equation}
which results in a de-projected three-dimensional density of
\begin{equation}
\rho_{{\rm pl}}(r) =  \frac{\rho_0}{\left[1+(r/r_{\rm
pl})^2\right]^{5/2}}.
\end{equation}
The only relevant free parameter in the Plummer profile is $r_{{\rm
pl}}$.

In Table~\ref{tab:parameterstable}  we show the respective fits to the
surface density for each of the dwarf
satellites~\cite{Irwin1995,SimonandGeha}. Many of the well-known dSphs
are well-fit by King profiles,  although for some galaxies the King
profile fits have been  updated to account for the observed
distribution of stars  in the outer regions~\cite{Munoz2006}.  As is
seen, the majority of the new satellites are well-fit  by Plummer
profiles.  In the instances where both Plummer  and King profiles have
been fit to the data, we find that the  exact form of the fit does not
strongly  affect the results we present below.

\subsection{Velocity Anisotropy}
We assume that both tangential components of the velocity  dispersion
are equal, $\sigma_\theta^2 = \sigma_\phi^2$.  Many of the systems we
study are observed to have  multiple stellar
populations~\cite{Mcconnachie2007},  so there is no reason why the
anisotropy should be constant throughout the galaxy. To account for
radial variation, we  parametrize the anisotropy profile as
\begin{equation}
\beta (r) =  (\beta_\infty-\beta_0) \frac{r^2}{r_\beta^2 + r^2} +
\beta_0.
\label{eq:betaprofile}
\end{equation}
The velocity anisotropy profile of this form is thus described by  an
asymptotic inner value,  $\beta_0$, and asymptotic value near the edge
of the halo,  $\beta_\infty$, and a scale radius, $r_\beta$. We  place
the constraints on $\beta_0$ and $\beta_\infty$  such that $\beta(r)
<1$ for all radii.

\subsection{Dark Matter Density Profile}
To model the dark matter mass distribution, we use a  density profile
of the form
\begin{equation} 
\rho(r) =  \frac{\rho_0}{(r/r_0)^a [1+(r/r_0)^b]^{(c-a)/b}}.
\label{eq:rhor}
\end{equation}
The asymptotic inner slope is determined by $a$, the  asymptotic outer
slope is determined by $c$, and the  transition between these two
regimes is determined by  $b$. The scale density is defined as
$\rho_0$, and the  scale radius is defined as $r_0$. In all cases, the
line-of-sight velocities are not able to determine the parameters
$a,b,c$, though as we see below, the data  do strongly constrain the
mass and the density of these  systems at a characteristic
radius~\cite{StrigariRedefining}.

For some regions of parameter space, the density profile  in
equation~\ref{eq:rhor} has an infinite mass, so it cannot  represent a
physically reasonable dark matter halo. To  ensure that the mass is
finite in all of parameter  space, we weight equation~\ref{eq:rhor} as
$\rho(r) \rightarrow \rho(r) \exp^{-r/r_{\rm cut}}$. We  determine the
cut-off radius, $r_{\rm cut}$, from the standard  Roche-limit criteria,
\begin{equation}
r_{\rm cut} \simeq  \left (\frac{G M_{\rm halo} D^2}{2
\sigma_{MW}^2}\right)^{1/3},
\label{eq:roche}
\end{equation} 
where $M_{\rm halo}$ is the total mass of the satellite,  $\sigma_{\rm
MW}$ is the velocity dispersion of the Milky  Way at the position of
the satellite, and $D$ is the distance  to the center of the
satellite. To establish a conservative  upper limit on $r_{\rm cut}$,
in particular allowing it to span the largest physically possible
range,  for all of the satellites  we take $M_{\rm halo} =10^9$
M$_\odot$ and $\sigma_{\rm MW} = 200$ km s$^{-1}$.

\section{Data Modeling}
\label{sec:likelihoodsection}
From a data set consisting of line-of-sight velocities,  our goal is
to determine the constraints on any of the parameters in
equation~\ref{eq:sigmaLOS}.  To evaluate the constraints on these
parameters, we undertake a maximum likelihood analysis and construct
the probability distribution for obtaining a set of line-of -sight
velocities in a dwarf galaxy.   The goal of this section is to present
and discuss our implementation of the maximum likelihood analysis.

\subsection{Likelihood Function}
The dispersion in the velocity at a given position is the sum of two
components:  1) the dispersion from the velocity distribution
function, and 2) the random error stemming from the uncertainty in the
measurement  of the velocity of each star. Motivated by  theoretical
modeling~\cite{Klimentowski:2006qe}  and observations of line-of-sight
velocities ~\cite{SimonandGeha}, we take both of these distributions
to be Gaussian.  The dispersion stemming from the velocity
distribution function is determined by the set of parameters in
equations~\ref{eq:betaprofile} and~\ref{eq:rhor} that govern the dark
matter halo model and the velocity anisotropy. We refer to this vector
set of parameters as $\vec{\theta}$.  The probability  for obtaining a
set of line-of-sight velocities, ${\bf v}$, given the theoretical
parameters, is $P({\bf v} | u,{\bf \sigma_t})$.  Here $u$ is defined
as the systemic velocity of the  galaxy, and ${\bf \sigma_t}$ is
determined  from equation~\ref{eq:sigmaLOS}.  It is thus implied that
${\bf \sigma_t}$ is a function of the parameters $\vec{\theta}$.  With
the above assumptions, the result for  $P({\bf v} | u,{\bf \sigma_t})$
is a Gaussian  distribution,
\begin{equation}
P({\bf v} | u,{\bf \sigma_t}) = \prod_{i=1}^N \frac{1}
{\sqrt{2\pi\sigma_\imath^2}} \exp\left[-\frac{1}{2}
\frac{(v_i-u)^2}{\sigma_\imath^2}\right].
\label{eq:likefull}
\end{equation}

The product is over the $N$ number of stars in the galaxy with
line-of-sight velocity  measurements.  The total variance at the
projected radius $R$ (at the position of the $i^{th}$ star) is thus
given by $\sigma_\imath^2 = \st^2 + \si^2$, where  $\st^2$ refers to
the variance from  the theoretical distribution, and $\si^2$ is the
variance  from the measurement uncertainty.  In writing
equation~\ref{eq:likefull} we have assumed no correlations between
both the theory and measured dispersions. Since we are assuming  no
streaming motion, $\st$ is replaced with equation~\ref{eq:sigmaLOS}.
According to Bayes theorem, equation~\ref{eq:likefull}  is
proportional to the probability of the parameters given the data,
i.e. $P({\bf v} | u,{\bf \sigma_t}) \propto P(u,{\bf \sigma_t}| {\bf
v})$.  When considered as a function of the parameters,
equation~\ref{eq:likefull} can be then defined as the likelihood
function for the parameters,  ${\cal L}(\vec{\theta})$.

It is important to note that the assumption of Gaussianity for the
intrinsic velocity distribution function is only  an approximation to
the true phase space distribution.  The true intrinsic distribution
function will depend  on the exact form of the density of the stellar
population and the potential of the background dark  matter
halo. Taking the stellar distribution to be a  King profile, the
Gaussian approximation provides a  good estimation of the true
distribution,  though there may be some deviations from Gaussianity in
the outer most regions of  galaxies~\cite{Walker2006}. From an
observational  perspective, the distribution  in velocities for nearly
all of the dwarf satellites we consider is  well-described by a
Gaussian distribution at the $\sim 2-3 \sigma$ level.   For all
satellites, we have found that tests of deviation from Gaussianity in
the data have  proved inconclusive.

In addition to deviations due to the nature of the  local
gravitational potential, from a systematic  perspective the observed
velocity distribution may  deviate from Gaussianity if there is large
contamination  to the sample from interloping stars or from stars not
bound  to the galaxy because of tidal interactions.  To accurately
model the  mass distribution of a galaxy with equations~\ref{eq:jeans}
and  ~\ref{eq:sigmaLOS} and line-of-sight velocities,  interloping and
unbound stars must be properly identified.  Positional and photometric
criteria provide a means for identifying these interloping and unbound
stars~\cite{SimonandGeha}. Numerical simulations have  also been
performed to aid in the identification of contaminating
stars~\cite{Piatek95,Klimentowski:2006qe}.  Tagging stars as outliers
based simply on their presence  in the tails of the velocity
distribution introduces an  intrinsic bias into the determination of
the mass, so  excluding these stars as non-members must be done with
caution. Once interlopers are cut, it is observed that  the mass
determined from the full-data sets is accurate  to $\sim
20\%$~\cite{Piatek95,Klimentowski:2006qe}.  This error is most
relevant, however, in determining the {\em total} halo mass. The
presence of interlopers is even less  significant when determining the
mass at radii nearer to the center of the halo, as we do below.

\section{Results} 
\label{sec:results} 
We now turn to application of the analysis above  and determine the
best fitting masses for the  Milky Way dwarf satellites. The primary
result will be integrated masses within radii of 0.1 and  0.3 kpc. The
latter provide the main results  of our analysis. The former is
convenient because, for a few satellites, it requires less of an
extrapolation of the dark matter halo  beyond the observed stellar
distribution.  Our choice of 0.3 kpc as the characteristic  radius
probes the mass further in the interior of the haloes  than the
characteristic radius of 0.6 kpc used in a previous
study~\cite{StrigariRedefining}.  This latter choice of 0.6 kpc was
better suited for comparing the masses of only the pre-SDSS, classical
dSphs, which are on  average more extended than the new SDSS
population.  For dwarf satellites with over a hundred line-of-sight
velocities,  we find the strength of the constraints on the mass
within 0.3 and 0.6 kpc to be typically similar.
 
The mass within a given radius, $m$, is obtained by  integrating the
probability distribution in  equation~\ref{eq:likefull} over the model
parameters,
\begin{equation} 
{\cal L}(m) \propto \int P[{\bf v} | u,{\bf \sigma_t}(\vec{\theta})]
\delta (m-M) d \vec{\theta}.
\label{eq:Lmass}
\end{equation}
In practice, determining ${\cal L}(m)$  then requires an integration
over all of the free parameters in equations~\ref{eq:betaprofile}
and~\ref{eq:rhor}. In
equation~\ref{eq:Lmass}, $M$ is determined by the halo model
parameters at the given point in parameter space.   Note also that in
equation~\ref{eq:Lmass} we have ignored the  normalization of the
likelihood,  which does not depend on the model parameters. To  ease
in quoting confidence limits, we simply normalize the peak of the
resulting integrated likelihood function to unity.

We assume uniform priors on the model parameters over their respective
ranges. We have tested  the assumption of uniform priors by also
considering Gaussian  priors on each parameter over their respective
allowed ranges.  For either assumed prior, our mass constraints are
found to be robust.  For the dark matter halo parameters, uniform
priors are chosen over the following ranges: $0.01< r_0 < 10$ kpc, $0
< a < 1.5$, $0.5 < b < 1.5$, and $2<c<4$. We find that all of these
parameters are not well-constrained by the line-of-sight velocity
data.  We also include a cut-off radius in the halo profile  as in
equation~\ref{eq:roche}, which, as discussed above, serves  to makes
the truncation of the halo sharper than the power law behavior implied
by the outer slope $c$.

For the radially-variable anisotropy model,  the following ranges are
taken:  $0.1 < r_\beta < 10$,  $-10 < \beta_\infty < 1$, and $-10 <
\beta_0 < 1$.   The choice of a flat prior on the velocity anisotropy
parameters  does in principle give a stronger weight to more
tangential orbits,  though we find that  including a prior on the
anisotropy that more uniformly weighs radial  orbits does not affect
the results that we present. Though we have assumed a
radially-variable anisotropy profile in equation~\ref{eq:betaprofile},
we find that our resulting constraints on the mass are equivalent to
those obtained in   a model where the velocity anisotropy is constant
in radius. The only instance in which our mass constraints weaken 
significantly is in the physically unlikely scenario that
the anisotropy profile has a  very steep transition right  around the
King (or Plummer) radius.

\begin{figure}
%\centering
\centerline{\psfig{figure=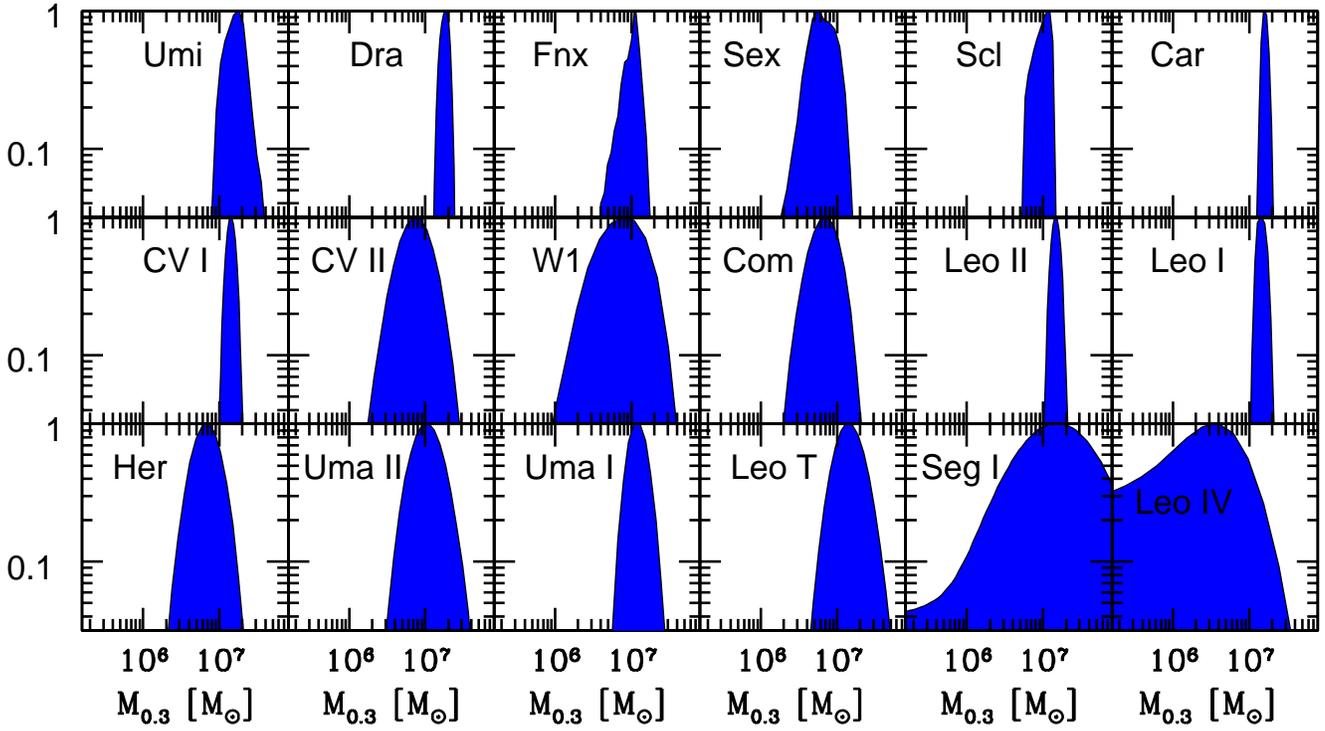,height=10cm}}
\caption{The likelihood function for the integrated mass within 0.3
  kpc for 18 of the Milky Way satellites. We marginalize over all
  parameters as described in the text.
\label{fig:like}
}
\end{figure}

Figure~\ref{fig:like} depicts the main results of  our analysis: the
mass likelihood functions  within 0.3 kpc for 18 of the Milky Way
dwarfs.  The classical, pre-SDSS dwarfs, which have more  measured
line-of-sight velocities spread over  a larger radial distance, have
likelihoods  that are much more strongly
constrained. Figure~\ref{fig:like} shows that the M$_{0.3}$ values are
tightly bunched around $\sim 10^7$ M$_\odot$. The only dwarf that
displays a large tail in its likelihood  at $\sim$ 1-$\sigma$ to small
M$_{0.3}$ is Leo IV, which may be related to the  shorter integration
time spent on Leo IV as well as the relatively small  sample of
stellar velocities in this galaxy.

The stellar luminosities for the Milky Way dwarf satellites   span a
range of approximately four orders of magnitude: from the  most
luminous satellite (Fornax, $\sim 10^7$ L$_\odot$), to the least
luminous  (Willman 1, Ursa Major II, Coma Berenices, and Segue 1:
$\sim 10^3$ L$_\odot$).  Since typical stellar mass-to-light ratios
are of order unity,  we can immediately deduce that all of these
dwarfs are strongly  dark matter dominated within the limiting radius
of their stellar distributions. Nearly all of these dwarfs  have
mass-to-light ratios within their inner 0.3 kpc of M$_{0.3}$/L$_\odot$
$> 10$.  The most luminous systems, including  Fornax, Sculptor, and
Leo I, have the smallest total M$_{0.3}$/L$_\odot$.  Because our
analysis determines the total dynamical mass, the  stars in these
galaxies likely make a significant contribution to the measured
M$_{0.3}$. However,  when casting the mass-to-light  ratio in terms of
total halo mass, M$_{\rm total}$, even for these most   luminous
dwarfs we have M$_{\rm total}/L > 100$~\cite{StrigariRedefining}.

In Table~\ref{tab:parameterstable}, we show the numerical results  for
the mass within 0.1 and 0.3 kpc for each dwarf satellite,  along with
other various observed properties. The results of
Table~\ref{tab:parameterstable} also show that the masses within 100
pc are well-constrained, implying a common mass of $\sim 10^6$
M$_\odot$ within this radius.   As mentioned above, for several dwarf
satellites, in particular Segue 1, Willman 1, and Coma Berenices,
determining the mass within 0.1 kpc requires less of an extrapolation
beyond the limiting radius of the stellar distributions. In  these
instances, the mass within 0.1 kpc is in good agreement  with masses
determined from a single-component, mass-follows-light
analysis~\cite{Illingworth1976,SimonandGeha,Munoz2006,Martin2007}.

Of the newly-discovered satellites, we have not displayed  results for
either Bootes I or Bootes II. Line-of-sight velocities for Bootes I
have been published~\cite{Munoz2006,Martin2007}, though  for ease of
comparison we have chosen to use common data sets. Our initial
estimate does in fact also show that  Bootes I is consistent with the
M$_{0.3} \simeq 10^7$ M$_\odot$ mass scale at $\sim 2-\sigma$. We are
currently unable to analyze Bootes II because it was discovered within
the past year  ~\cite{Walsh2007} and does not yet have measured
line-of-sight  velocities. Of the  classical, pre-SDSS satellites, we
do not determine the mass for the Large Magellanic Cloud (LMC), Small
Magellanic Cloud (SMC), or Sagittarius. It is likely that both the LMC
and SMC have significant baryonic contribution to the their respective
masses within 0.3 kpc, so a detailed determination of the halo mass of
each galaxy will require accurate modeling of this
contribution. Sagittarius has published line-of-sight velocities
~\cite{Ibata1997}, though because it is in the process of  tidal
disruption, streaming motions will strongly affect the  determination
of its mass, making the above analysis not  reliable.

As discussed above, for the respective stellar  distributions, we use
the King or Plummer profiles given in
Table~\ref{tab:parameterstable}. For several of the classical dwarfs,
it has been found that in the outer regions of the galaxy  a faint
population of stars exists that has a surface density that falls off
less  steeply than either the King or Plummer profile
~\cite{Majewski2000,Munoz2006}.  Interpreting these stars as bound to
the galaxy would require an increase in the {\rm total}  halo
mass. However, we find that the  presence of these stars has little
effect on the mass within 0.3 kpc,  as this latter quantity is
primarily  determined by the population of stars that are
well-described by the respective   parameters in
Table~\ref{tab:parameterstable}.

It is remarkable to note that, when considering both the wide  range
of scale radii in Table~\ref{tab:parameterstable} and the range of
velocity dispersions, the common mass scale persists. Though
determination of the masses requires  a numerical solution to the
Jeans equation as described above, we can gain some insight into the
scaling of the mass with both the velocity  dispersion and the scale
radius by considering the limit in which $\beta = 0$, the surface
density of stars is given by a Plummer  profile, and the velocity
dispersion is constant as a function of radius. The latter
simplification is a good approximation for many of the dwarfs. Under
these assumptions, the mass at any radius is given by $M(r) \simeq 1.2
\times 10^3$ M$_\odot$ $\left[\frac{r}{\rm
pc}\right]\left[\frac{\sigma}{\rm km/s}\right]^2 \frac{y^2}{1+y^2}$,
where $y=r/r_{pl}$. Examining this formula, we see that, for a fixed
projected velocity dispersion, the  mass will {\em increase} if the
Plummer scale radius is reduced.  Qualitatively, this scaling can be
understood by noting that it takes a deeper potential well to confine
stars with a larger dispersion to a small radius. We note that all of
our results are consistent with the minimum mass obtained from the
virial  theorem~\cite{Merritt1987}; for example for Willman 1 we
obtain  a minimum mass of $\sim 3 \times 10^5$ M$_\odot$, and for
Fornax we obtain a minimum mass of $\sim 2 \times 10^6$ M$_\odot$.

It should be stressed again that in our dynamical mass
analysis we have assumed that the stars are
orbiting in a spherically symmetric potential. This is a reasonable
assumption because numerical simulations show~\cite{Kuhlen2007} that
potentials of these dark matter haloes are close to spherical with
triaxial axis ratios of 0.8 to 0.9. Dispersion and bias introduced in
the dynamical mass determinations at the level of 10-20\% are not
important at the present stage but will be as data sets get larger for
the fainter satellites.

\section{Tidal Disruption, Rotation, and Binary Contamination}
Equation~\ref{eq:likefull} assumes that rotational motion  or external
tidal forces do not contribute to the observed  line-of-sight
velocities. However, this may not be a complete description of the
dwarf satellites, as it is very likely that their dark matter  haloes have 
been affected by tidal interactions in the past.  Although the simple
scalings above indicate that it is unlikely  tidal forces dominate the
dynamics of these systems, it is important  to quantify any deviations
from steady-state dynamics more precisely. The goal of this section is to 
use the line-of-sight
velocity data  to provide conservative tests for both rotation and
tidal disruption.

\subsection{Tidal Disruption} 
Tidal forces induce a gradient in the line-of-sight 
velocities across a galaxy~\cite{Piatek95,Fellhauer:2006jr}. 
To test for a velocity gradient, we describe the
stars in terms of their projected radial distance, $R$, from the
center of the dwarf,  and the azimuthal angle of each star, $\phi$,
relative to a fixed coordinate system. Streaming motion would
thus appear as a velocity gradient about an angle $\phi_0$, which is
determined from the data  ~\cite{Drukier1998}. In the presence of
streaming,  the systemic velocity in the likelihood function of
equation~\ref{eq:likefull} is replaced as
\begin{equation}
u \rightarrow u + A\sin(\phi_i + \phi_0).
\label{eq:rotation_sin}
\end{equation}
Here $A$ represents the  amplitude of the streaming motion,  $\phi_0$
is the (projected) axis of the streaming motion, and $\phi_\imath$ is
the azimuthal angle of the $\imath^{th}$  star. It is important to note that, 
in principle, replacing the
systemic velocity with equation~\ref{eq:rotation_sin} provides only an
approximation to streaming motion  that serves to pick out a dipolar
term in the velocity field.  Higher order multipoles may exist as a
result of tidal forces; however for our purposes we assume these
higher order terms to be sub-dominant to the leading dipolar term
given in  equation~\ref{eq:rotation_sin}.

\begin{figure}
%\centering
\centerline{\psfig{figure=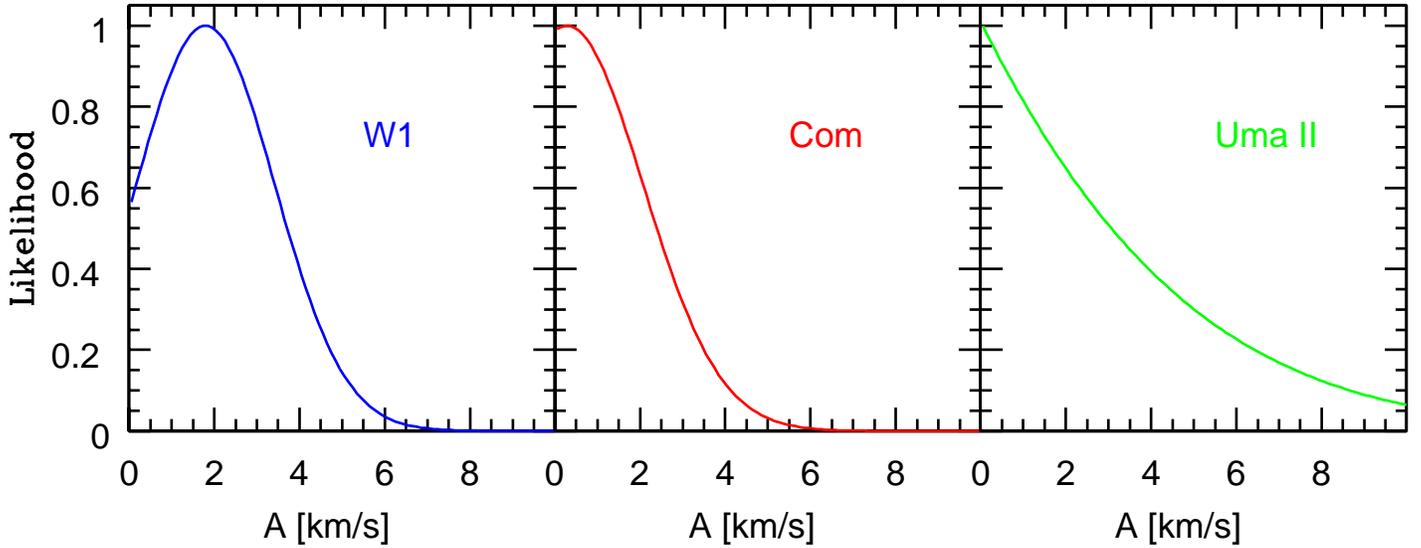}}
\caption{The likelihood function for the rotation amplitude for
Willman 1, Coma Berenices, and Ursa Major II.
\label{fig:Arot}
}
\end{figure}

A non-zero detection of $A$ suggests the presence of tidal
forces or rotational motion. We have examined the line-of-sight velocities
for each system  and have determined the likelihood for the amplitude
$A$ after marginalizing over all of the halo parameters, anisotropy
parameters, as well as $\phi_0$.  In Figure~\ref{fig:Arot}, we show
the resulting likelihood functions for $A$ for three systems: Willman
1, Coma Berenics, and Ursa Major II. We show these systems as examples
because they are the amongst the nearest to the Milky Way, and thus
may be plausible candidates for tidal disruption.  As is seen, there
is no significant detection of $A$.  For the Willman 1 sample we use, the
mean value of $A$ is $\sim 2$ km s$^{-1}$,  but this object is 
still consistent with no rotation at $\sim 1-\sigma$.

In all of the remaining newly-discovered satellites, we find no statistically
significant detection of $A$. In nearly all cases, we place strong
upper limits on $A$ to be a fraction of the intrinsic dispersion of
each system. We note that the inconclusive detection of velocity
gradients in Coma Berenices and Ursa Major II differ from the  results
reported in Simon and Geha~\cite{SimonandGeha}, where a weighted mean
of the velocities on each side of the position angle showed an
apparent velocity gradient. Here we find that accounting for
correlations with higher order velocity moments washes out the
detection of a velocity gradient in each of these systems. These
results imply that rotational motion or tidal disruption  in the
above parametrization likely does not  significantly affect the 
mass modeling. 

We note that, when interpreting the parametrization  in
equation~\ref{eq:rotation_sin} as a rotational signal,  it is only an
approximation to the true three-dimensional rotation. To
construct a more physically-plausible model  one would need to replace
equation~\ref{eq:rotation_sin} with a model that depends on at least
two angles that describe both the inclination of the rotation axis and
the line-of-nodes of the system.  Additionally, flattening of the stellar
distribution must be accounted for by modifying the spherically-symmetric
Jeans equation. At the very least, these effects will increase the uncertainty on
the rotation  amplitude,  so in  this sense our errors on the rotation
amplitude are likely too strict.  Our initial estimates show that, when adding 
an additional angle to account for three-dimensional rotation, the error on the
amplitude of the rotational motion increases by a factor of about two.

\subsection{Perspective Rotation} As
the satellites are extended objects on the sky, the line-of-sight
velocity will vary as we move across the object. As we move farther
from the center of the system, the tangential motion of the object
will contribute to the measured line-of-sight velocity. This effect is
known as ``perspective rotation,''  and the net result is that a
non-rotating object will appear to have a velocity gradient
across the system~\cite{Feast1961}.

Perspective rotation can be simply parametrized by
replacing the systemic velocity in the likelihood function as
%\begin{equation}
$u \rightarrow u + v_x x/D + v_y y/D$,
%\end{equation} 
where $x$ and $y$ are the projected positions on the sky and  $v_x$
and $v_y$ are the tangential motions in these directions.  For the
purposes of our present analysis, we are interested in  determining if
the addition of the perspective rotation term  will alter our mass
estimates derived above. We find that, in all data sets, the masses
we determine are robust even after accounting for perspective rotation. 
A full analysis of a given data set 
will require  addition of both the streaming motion term in
equation~\ref{eq:rotation_sin} and the perspective rotation term;
adding both of these  will result in further degeneracies that will
make  each separate effect more difficult to extract.

\subsection{Contamination from Binary Stars}

A final effect we cannot include at present that may introduce  a
systematic in determining the masses is the
contribution of internal binary star motion to the velocity
dispersion. For binaries with orbits of order 100 AU, the relative
binary speed is comparable to the intrinsic velocity dispersion of the
satellite galaxy. Thus the bias due to binaries depends on the
fraction of stars that are binaries with orbits smaller than about 100
AU for the stellar population whose velocities are being measured. For
two of the most luminous satellites, Draco and Ursa Minor (which also
have a higher velocity dispersion), a significant effect of internal
binary motion on the measured velocity dispersion has been shown to be
unlikely~\cite{Olszewski1996}. A detailed study for faint dwarfs
will require repeat observations of stars a year or two apart; this
should be available in the near future. Here we simply note that if a
significant fraction of the measured dispersion was due to binary
motion, then the distribution of velocities would deviate
significantly from the observed Gaussian distribution.

%\bibliography{supp1} \setlength{\tabcolsep}{0.25cm}
\clearpage
\begin{table}
%\begin{center}
%{\bf Properties of Milky Way Satellites\vspace*{2mm}}
{\scriptsize
\begin{tabular}{lccccccccc}
\hline Satellite & Distance [kpc] & $r_{king}$, $r_{pl}$
[kpc]&$r_{lim}$ [kpc]  & Luminosity [10$^6$ L$_\odot$]  & M$_{0.1}$
[10$^7$ M$_\odot$]  & M$_{0.3}$ [10$^7$ M$_\odot$]\\ \hline \hline
Ursa Minor (Umi)& 66& 0.30& 1.50 & $0.29$& $0.21_{-0.14}^{+0.09}$ &$1.79_{-0.59}^{+0.37}$\\ 
Draco (Dra)& 80 & 0.18 & 0.93 & $0.26$ &$0.09_{-0.02}^{+0.20}$&$1.87_{-0.29}^{+0.20}$\\ 
Sculptor (Scl) & 80& 0.28& 1.63 & $2.15$&$0.15_{-0.10}^{+0.28}$& $1.20_{-0.37}^{+0.11}$\\ 
Sextans (Sex) &86 &0.40 & 4.00 &  $0.50$  &$0.06_{-0.01}^{+0.02}$& $0.57_{-0.14}^{+0.45}$\\
Carina (Car) &101&0.26 & 0.85  & $0.43$ & $0.48_{-0.06}^{+0.07}$&$1.57_{-0.10}^{+0.19}$\\ 
Fornax (Fnx) &138 &0.39 & 2.70 & $15.5$ & $0.12_{-0.04}^{+0.07}$& $1.14_{-0.12}^{+0.09}$\\
Leo II &205 & 0.19 & 0.52 & 0.58  &$0.16_{-0.07}^{+0.03}$&$1.43_{-0.15}^{+0.23}$\\ 
Leo I &250  & 0.20  & 0.80 & 4.79 &$0.06_{-0.01}^{+0.14}$ & $1.45_{-0.20}^{+0.27}$\\ 
\hline  
Segue 1 (Seg 1) & 25 & 0.031 & -- & $3.4 \times 10^{-4}$ &$0.35_{-0.24}^{+0.58}$& $1.58_{-1.11}^{+3.30}$\\ 
Ursa Major II (Uma II) & 32 & 0.127 &-- & $4.0 \times 10^{-3}$&$0.31_{-0.10}^{+0.18}$&$1.09_{-0.44}^{+0.89}$\\ 
Willman 1 (W1) & 38 &0.025 & -- & $1.0 \times 10^{-3}$ &$0.23_{-0.09}^{+0.18}$&$0.77_{-0.42}^{+0.89}$\\ 
Coma (Com) & 44 &0.064 & -- & $3.7 \times 10^{-3}$  &$0.19_{-0.05}^{+0.09}$&$0.72_{-0.28}^{+0.36}$\\ 
Ursa Major I (Uma I) &106 &0.308 & -- & $1.4 \times 10^{-2}$ &$0.34_{-0.09}^{+0.15}$&$1.10_{-0.29}^{+0.70}$\\ 
Hercules (Her) & 138&0.321 & -- & $3.6 \times 10^{-2}$& $0.19_{-0.07}^{+0.10}$&$0.72_{-0.21}^{+0.51}$\\ 
CV II & 151& 0.132 & -- & $7.9 \times10^{-3}$ &$0.19_{-0.07}^{+0.14}$ & $0.70_{-0.25}^{+0.53}$\\ 
Leo IV & 158 & 0.152 & -- & $8.7 \times 10^{-3}$ &$0.12_{-0.09}^{+0.14}$ & $0.39_{-0.29}^{+0.50}$\\ 
CV I & 224& 0.554 &-- & $2.3 \times 10^{-1}$ &$0.34_{-0.08}^{+0.20}$&$1.40_{-0.19}^{+0.18}$\\ 
Leo T & 417 &0.170 & -- & $5.9 \times10^{-2}$ &$0.39_{-0.13}^{+0.25}$& $1.30_{-0.42}^{+0.88}$\\ \hline
\hline
\end{tabular}
%\end{center}
\caption{{\scriptsize Table of properties of Milky Way satellites.
Where shown, in parenthesis we list the abbreviation used in
Figure~\ref{fig:like}. Dwarf satellites listed below the  horizontal
dividing line  denote the newly-discovered population of SDSS
dwarfs. For the classical satellites, we use King profiles, described
by the King core radius, $r_{king}$. For the new satellites, we use
Plummer profiles,  described by $r_{pl}$. The limiting radius,
$r_{lim}$, is  defined as the limiting radius for the King profile.
The error bars on the mass values reflect the points at which the
likelihood  function falls off to 60.6\% of its peak value on either
side. The luminosities for the newly-discovered satellites are 
taken from Ref.~\cite{Martin:2008wj}.}
\label{tab:parameterstable}} }
\end{table}

\clearpage 

\clearpage 

%\end{document}

\bibliography{letter}

\begin{thebibliography}{10}
\expandafter\ifx\csname url\endcsname\relax
  \def\url#1{\texttt{#1}}\fi
\expandafter\ifx\csname urlprefix\endcsname\relax\def\urlprefix{URL }\fi
\providecommand{\bibinfo}[2]{#2}
\providecommand{\eprint}[2][]{\url{#2}}

\bibitem{Willman2005}
\bibinfo{author}{{Willman}, B.} \emph{et~al.}
\newblock \bibinfo{title}{{A New Milky Way Companion: Unusual Globular Cluster
  or Extreme Dwarf Satellite?}}
\newblock \emph{\bibinfo{journal}{Astron. J.}} \textbf{\bibinfo{volume}{129}},
  \bibinfo{pages}{2692--2700} (\bibinfo{year}{2005}).

\bibitem{Belokurov2007}
\bibinfo{author}{{Belokurov}, V.} \emph{et~al.}
\newblock \bibinfo{title}{{Cats and Dogs, Hair and a Hero: A Quintet of New
  Milky Way Companions}}.
\newblock \emph{\bibinfo{journal}{Astrophys. J.}}
  \textbf{\bibinfo{volume}{654}}, \bibinfo{pages}{897--906}
  (\bibinfo{year}{2007}).

\bibitem{Mateo1998}
\bibinfo{author}{{Mateo}, M.~L.}
\newblock \bibinfo{title}{{Dwarf Galaxies of the Local Group}}.
\newblock \emph{\bibinfo{journal}{Ann. Rev. Astron. Astrophys.}}
  \textbf{\bibinfo{volume}{36}}, \bibinfo{pages}{435--506}
  (\bibinfo{year}{1998}).

\bibitem{Gilmore2007}
\bibinfo{author}{{Gilmore}, G.} \emph{et~al.}
\newblock \bibinfo{title}{{The Observed Properties of Dark Matter on Small
  Spatial Scales}}.
\newblock \emph{\bibinfo{journal}{Astrophys. J.}}
  \textbf{\bibinfo{volume}{663}}, \bibinfo{pages}{948--959}
  (\bibinfo{year}{2007}).

\bibitem{Spergel2007}
\bibinfo{author}{{Spergel}, D.~N.} \emph{et~al.}
\newblock \bibinfo{title}{{Three-Year Wilkinson Microwave Anisotropy Probe
  (WMAP) Observations: Implications for Cosmology}}.
\newblock \emph{\bibinfo{journal}{Astrophys. J.}}
  \textbf{\bibinfo{volume}{170}}, \bibinfo{pages}{S377--S408}
  (\bibinfo{year}{2007}).

\bibitem{SimonandGeha}
\bibinfo{author}{{Simon}, J.~D.} \& \bibinfo{author}{{Geha}, M.}
\newblock \bibinfo{title}{{The Kinematics of the Ultra-faint Milky Way
  Satellites: Solving the Missing Satellite Problem}}.
\newblock \emph{\bibinfo{journal}{Astrophys. J.}}
  \textbf{\bibinfo{volume}{670}}, \bibinfo{pages}{313--331}
  (\bibinfo{year}{2007}).

\bibitem{Walker2007}
\bibinfo{author}{{Walker}, M.~G.} \emph{et~al.}
\newblock \bibinfo{title}{{Velocity Dispersion Profiles of Seven Dwarf
  Spheroidal Galaxies}}.
\newblock \emph{\bibinfo{journal}{Astrophys. J.}}
  \textbf{\bibinfo{volume}{667}}, \bibinfo{pages}{L53--L56}
  (\bibinfo{year}{2007}).

\bibitem{Peebles1982}
\bibinfo{author}{{Peebles}, P.~J.~E.}
\newblock \bibinfo{title}{{Large-scale background temperature and mass
  fluctuations due to scale-invariant primeval perturbations}}.
\newblock \emph{\bibinfo{journal}{Astrophys. J.}}
  \textbf{\bibinfo{volume}{263}}, \bibinfo{pages}{L1--L5}
  (\bibinfo{year}{1982}).

\bibitem{White1983}
\bibinfo{author}{{White}, S.~D.~M.}, \bibinfo{author}{{Frenk}, C.~S.} \&
  \bibinfo{author}{{Davis}, M.}
\newblock \bibinfo{title}{{Clustering in a neutrino-dominated universe}}.
\newblock \emph{\bibinfo{journal}{Astrophys. J.}}
  \textbf{\bibinfo{volume}{274}}, \bibinfo{pages}{L1--L5}
  (\bibinfo{year}{1983}).

\bibitem{Blumenthal1984}
\bibinfo{author}{{Blumenthal}, G.~R.}, \bibinfo{author}{{Faber}, S.~M.},
  \bibinfo{author}{{Primack}, J.~R.} \& \bibinfo{author}{{Rees}, M.~J.}
\newblock \bibinfo{title}{{Formation of galaxies and large-scale structure with
  cold dark matter}}.
\newblock \emph{\bibinfo{journal}{Nature}} \textbf{\bibinfo{volume}{311}},
  \bibinfo{pages}{517--525} (\bibinfo{year}{1984}).

\bibitem{Klypin1999}
\bibinfo{author}{{Klypin}, A.}, \bibinfo{author}{{Kravtsov}, A.~V.},
  \bibinfo{author}{{Valenzuela}, O.} \& \bibinfo{author}{{Prada}, F.}
\newblock \bibinfo{title}{{Where Are the Missing Galactic Satellites?}}
\newblock \emph{\bibinfo{journal}{Astrophys. J.}}
  \textbf{\bibinfo{volume}{522}}, \bibinfo{pages}{82--92}
  (\bibinfo{year}{1999}).

\bibitem{Moore1999}
\bibinfo{author}{{Moore}, B.} \emph{et~al.}
\newblock \bibinfo{title}{{Dark Matter Substructure within Galactic Halos}}.
\newblock \emph{\bibinfo{journal}{Astrophys. J.}}
  \textbf{\bibinfo{volume}{524}}, \bibinfo{pages}{L19--L22}
  (\bibinfo{year}{1999}).

\bibitem{Diemand2005}
\bibinfo{author}{{Diemand}, J.}, \bibinfo{author}{{Moore}, B.} \&
  \bibinfo{author}{{Stadel}, J.}
\newblock \bibinfo{title}{{Earth-mass dark-matter haloes as the first
  structures in the early Universe}}.
\newblock \emph{\bibinfo{journal}{Nature}} \textbf{\bibinfo{volume}{433}},
  \bibinfo{pages}{389--391} (\bibinfo{year}{2005}).

\bibitem{Diemand:2006ik}
\bibinfo{author}{Diemand, J.}, \bibinfo{author}{Kuhlen, M.} \&
  \bibinfo{author}{Madau, P.}
\newblock \bibinfo{title}{{Dark matter substructure and gamma-ray annihilation
  in the Milky Way halo}}.
\newblock \emph{\bibinfo{journal}{Astrophys. J.}}
  \textbf{\bibinfo{volume}{657}}, \bibinfo{pages}{262} (\bibinfo{year}{2007}).

\bibitem{Bode:2000gq}
\bibinfo{author}{Bode, P.}, \bibinfo{author}{Ostriker, J.~P.} \&
  \bibinfo{author}{Turok, N.}
\newblock \bibinfo{title}{{Halo Formation in Warm Dark Matter Models}}.
\newblock \emph{\bibinfo{journal}{Astrophys. J.}}
  \textbf{\bibinfo{volume}{556}}, \bibinfo{pages}{93--107}
  (\bibinfo{year}{2001}).

\bibitem{Efstathiou1992}
\bibinfo{author}{{Efstathiou}, G.}
\newblock \bibinfo{title}{{Suppressing the formation of dwarf galaxies via
  photoionization}}.
\newblock \emph{\bibinfo{journal}{Mon. Not. R. Astron. Soc.}}
  \textbf{\bibinfo{volume}{256}}, \bibinfo{pages}{43P--47P}
  (\bibinfo{year}{1992}).

\bibitem{Kauffmann1993}
\bibinfo{author}{{Kauffmann}, G.}, \bibinfo{author}{{White}, S.~D.~M.} \&
  \bibinfo{author}{{Guiderdoni}, B.}
\newblock \bibinfo{title}{{The Formation and Evolution of Galaxies Within
  Merging Dark Matter Haloes}}.
\newblock \emph{\bibinfo{journal}{Mon. Not. R. Astron. Soc.}}
  \textbf{\bibinfo{volume}{264}}, \bibinfo{pages}{201--+}
  (\bibinfo{year}{1993}).

\bibitem{Bullock2001}
\bibinfo{author}{{Bullock}, J.~S.}, \bibinfo{author}{{Kravtsov}, A.~V.} \&
  \bibinfo{author}{{Weinberg}, D.~H.}
\newblock \bibinfo{title}{{Reionization and the Abundance of Galactic
  Satellites}}.
\newblock \emph{\bibinfo{journal}{Astrophys. J.}}
  \textbf{\bibinfo{volume}{539}}, \bibinfo{pages}{517--521}
  (\bibinfo{year}{2000}).

\bibitem{Kravtsov2004}
\bibinfo{author}{{Kravtsov}, A.~V.}, \bibinfo{author}{{Gnedin}, O.~Y.} \&
  \bibinfo{author}{{Klypin}, A.~A.}
\newblock \bibinfo{title}{{The Tumultuous Lives of Galactic Dwarfs and the
  Missing Satellites Problem}}.
\newblock \emph{\bibinfo{journal}{Astrophys. J.}}
  \textbf{\bibinfo{volume}{609}}, \bibinfo{pages}{482--497}
  (\bibinfo{year}{2004}).

\bibitem{Mayer2007}
\bibinfo{author}{{Mayer}, L.}, \bibinfo{author}{{Kazantzidis}, S.},
  \bibinfo{author}{{Mastropietro}, C.} \& \bibinfo{author}{{Wadsley}, J.}
\newblock \bibinfo{title}{{Early gas stripping as the origin of the darkest
  galaxies in the Universe}}.
\newblock \emph{\bibinfo{journal}{Nature}} \textbf{\bibinfo{volume}{445}},
  \bibinfo{pages}{738--740} (\bibinfo{year}{2007}).

\bibitem{Munoz2006}
\bibinfo{author}{{Mu{\~n}oz}, R.~R.} \emph{et~al.}
\newblock \bibinfo{title}{{Exploring Halo Substructure with Giant Stars: The
  Dynamics and Metallicity of the Dwarf Spheroidal in Bo{\"o}tes}}.
\newblock \emph{\bibinfo{journal}{Astrophys. J.}}
  \textbf{\bibinfo{volume}{650}}, \bibinfo{pages}{L51--L54}
  (\bibinfo{year}{2006}).

\bibitem{Martin2007}
\bibinfo{author}{{Martin}, N.~F.}, \bibinfo{author}{{Ibata}, R.~A.},
  \bibinfo{author}{{Chapman}, S.~C.}, \bibinfo{author}{{Irwin}, M.} \&
  \bibinfo{author}{{Lewis}, G.~F.}
\newblock \bibinfo{title}{{A Keck/DEIMOS spectroscopic survey of faint Galactic
  satellites: searching for the least massive dwarf galaxies}}.
\newblock \emph{\bibinfo{journal}{Mon. Not. R. Astron. Soc.}}
  \textbf{\bibinfo{volume}{380}}, \bibinfo{pages}{281--300}
  (\bibinfo{year}{2007}).

\bibitem{StrigariRedefining}
\bibinfo{author}{{Strigari}, L.~E.} \emph{et~al.}
\newblock \bibinfo{title}{{Redefining the Missing Satellites Problem}}.
\newblock \emph{\bibinfo{journal}{Astrophys. J.}}
  \textbf{\bibinfo{volume}{669}}, \bibinfo{pages}{676--683}
  (\bibinfo{year}{2007}).

\bibitem{Mateo1993}
\bibinfo{author}{{Mateo}, M.}, \bibinfo{author}{{Olszewski}, E.~W.},
  \bibinfo{author}{{Pryor}, C.}, \bibinfo{author}{{Welch}, D.~L.} \&
  \bibinfo{author}{{Fischer}, P.}
\newblock \bibinfo{title}{{The Carina dwarf spheroidal galaxy - How dark is
  it?}}
\newblock \emph{\bibinfo{journal}{Astron. J.}} \textbf{\bibinfo{volume}{105}},
  \bibinfo{pages}{510--526} (\bibinfo{year}{1993}).

\bibitem{Piatek95}
\bibinfo{author}{{Piatek}, S.} \& \bibinfo{author}{{Pryor}, C.}
\newblock \bibinfo{title}{{The effect of galactic tides on the apparent
  mass-to-light ratios in dwarf spheroidal galaxies}}.
\newblock \emph{\bibinfo{journal}{Astron. J.}} \textbf{\bibinfo{volume}{109}},
  \bibinfo{pages}{1071--1085} (\bibinfo{year}{1995}).

\bibitem{Fellhauer:2006jr}
\bibinfo{author}{Fellhauer, M.} \emph{et~al.}
\newblock \bibinfo{title}{{Is Ursa Major II the Progenitor of the Orphan
  Stream?}}
\newblock \emph{\bibinfo{journal}{Mon. Not. R. Astron. Soc.}}
  \textbf{\bibinfo{volume}{375}}, \bibinfo{pages}{1171--1179}
  (\bibinfo{year}{2007}).

\bibitem{Dunkley:2008ie}
\bibinfo{author}{Dunkley, J.} \emph{et~al.}
\newblock \bibinfo{title}{{Five-Year Wilkinson Microwave Anisotropy Probe
  (WMAP) Observations: Likelihoods and Parameters from the WMAP data}}
  (\bibinfo{year}{2008}).
\newblock \eprint{arXiv:0803.0586 [astro-ph]}.

\bibitem{BullockCvir}
\bibinfo{author}{{Bullock}, J.~S.} \emph{et~al.}
\newblock \bibinfo{title}{{Profiles of dark haloes: evolution, scatter and
  environment}}.
\newblock \emph{\bibinfo{journal}{Mon. Not. R. Astron. Soc.}}
  \textbf{\bibinfo{volume}{321}}, \bibinfo{pages}{559--575}
  (\bibinfo{year}{2001}).

\bibitem{DekelSilk}
\bibinfo{author}{{Dekel}, A.} \& \bibinfo{author}{{Silk}, J.}
\newblock \bibinfo{title}{{The origin of dwarf galaxies, cold dark matter, and
  biased galaxy formation}}.
\newblock \emph{\bibinfo{journal}{Astrophys. J.}}
  \textbf{\bibinfo{volume}{303}}, \bibinfo{pages}{39--55}
  (\bibinfo{year}{1986}).

\bibitem{Wyithe2006}
\bibinfo{author}{{Wyithe}, J.~S.~B.} \& \bibinfo{author}{{Loeb}, A.}
\newblock \bibinfo{title}{{Suppression of dwarf galaxy formation by cosmic
  reionization}}.
\newblock \emph{\bibinfo{journal}{Nature}} \textbf{\bibinfo{volume}{441}},
  \bibinfo{pages}{322--324} (\bibinfo{year}{2006}).

\bibitem{Walker2006}
\bibinfo{author}{{Walker}, M.~G.} \emph{et~al.}
\newblock \bibinfo{title}{{Internal Kinematics of the Fornax Dwarf Spheroidal
  Galaxy}}.
\newblock \emph{\bibinfo{journal}{Astron. J.}} \textbf{\bibinfo{volume}{131}},
  \bibinfo{pages}{2114--2139} (\bibinfo{year}{2006}).

\bibitem{Koch2007}
\bibinfo{author}{{Koch}, A.} \emph{et~al.}
\newblock \bibinfo{title}{{Stellar Kinematics in the Remote Leo II Dwarf
  Spheroidal Galaxy-Another Brick in the Wall}}.
\newblock \emph{\bibinfo{journal}{Astron. J.}} \textbf{\bibinfo{volume}{134}},
  \bibinfo{pages}{566--578} (\bibinfo{year}{2007}).

\bibitem{Kuhlen2007}
\bibinfo{author}{{Kuhlen}, M.}, \bibinfo{author}{{Diemand}, J.} \&
  \bibinfo{author}{{Madau}, P.}
\newblock \bibinfo{title}{{The Shapes, Orientation, and Alignment of Galactic
  Dark Matter Subhalos}}.
\newblock \emph{\bibinfo{journal}{Astrophys. J.}}
  \textbf{\bibinfo{volume}{671}}, \bibinfo{pages}{1135--1146}
  (\bibinfo{year}{2007}).

\bibitem{BinneyTremaine}
\bibinfo{author}{{Binney}, J.} \& \bibinfo{author}{{Tremaine}, S.}
\newblock \emph{\bibinfo{title}{{Galactic dynamics}}}
  (\bibinfo{publisher}{Princeton, NJ, Princeton University Press, 1987, 747
  p.}, \bibinfo{year}{1987}).

\bibitem{King1962}
\bibinfo{author}{King, I.}
\newblock \bibinfo{title}{The structure of star clusters. i. an empirical
  density law}.
\newblock \emph{\bibinfo{journal}{Astron. J.}} \textbf{\bibinfo{volume}{67}},
  \bibinfo{pages}{471} (\bibinfo{year}{1962}).

\bibitem{Irwin1995}
\bibinfo{author}{{Irwin}, M.} \& \bibinfo{author}{{Hatzidimitriou}, D.}
\newblock \bibinfo{title}{{Structural parameters for the Galactic dwarf
  spheroidals}}.
\newblock \emph{\bibinfo{journal}{Mon. Not. R. Astron. Soc.}}
  \textbf{\bibinfo{volume}{277}}, \bibinfo{pages}{1354--1378}
  (\bibinfo{year}{1995}).

\bibitem{Mcconnachie2007}
\bibinfo{author}{{McConnachie}, A.~W.}, \bibinfo{author}{{Pe{\~n}arrubia}, J.}
  \& \bibinfo{author}{{Navarro}, J.~F.}
\newblock \bibinfo{title}{{Multiple dynamical components in Local Group dwarf
  spheroidals}}.
\newblock \emph{\bibinfo{journal}{Mon. Not. R. Astron. Soc.}}
  \textbf{\bibinfo{volume}{380}}, \bibinfo{pages}{L75--L79}
  (\bibinfo{year}{2007}).

\bibitem{Klimentowski:2006qe}
\bibinfo{author}{Klimentowski, J.} \emph{et~al.}
\newblock \bibinfo{title}{Mass modelling of dwarf spheroidal galaxies: the
  effect of unbound stars from tidal tails and the milky way}.
\newblock \emph{\bibinfo{journal}{Mon. Not. R. Astron. Soc.}}
  \textbf{\bibinfo{volume}{378}}, \bibinfo{pages}{353--368}
  (\bibinfo{year}{2007}).

\bibitem{Illingworth1976}
\bibinfo{author}{{Illingworth}, G.}
\newblock \bibinfo{title}{{The masses of globular clusters. II - Velocity
  dispersions and mass-to-light ratios}}.
\newblock \emph{\bibinfo{journal}{Astrophys. J.}}
  \textbf{\bibinfo{volume}{204}}, \bibinfo{pages}{73--93}
  (\bibinfo{year}{1976}).

\bibitem{Walsh2007}
\bibinfo{author}{{Walsh}, S.~M.}, \bibinfo{author}{{Jerjen}, H.} \&
  \bibinfo{author}{{Willman}, B.}
\newblock \bibinfo{title}{{A Pair of Bo{\"o}tes: A New Milky Way Satellite}}.
\newblock \emph{\bibinfo{journal}{Astrophys. J.}}
  \textbf{\bibinfo{volume}{662}}, \bibinfo{pages}{L83--L86}
  (\bibinfo{year}{2007}).

\bibitem{Ibata1997}
\bibinfo{author}{{Ibata}, R.~A.}, \bibinfo{author}{{Wyse}, R.~F.~G.},
  \bibinfo{author}{{Gilmore}, G.}, \bibinfo{author}{{Irwin}, M.~J.} \&
  \bibinfo{author}{{Suntzeff}, N.~B.}
\newblock \bibinfo{title}{{The Kinematics, Orbit, and Survival of the
  Sagittarius Dwarf Spheroidal Galaxy}}.
\newblock \emph{\bibinfo{journal}{Astron. J.}} \textbf{\bibinfo{volume}{113}},
  \bibinfo{pages}{634--655} (\bibinfo{year}{1997}).

\bibitem{Majewski2000}
\bibinfo{author}{{Majewski}, S.~R.} \emph{et~al.}
\newblock \bibinfo{title}{{Exploring Halo Substructure with Giant Stars. II.
  Mapping the Extended Structure of the Carina Dwarf Spheroidal Galaxy}}.
\newblock \emph{\bibinfo{journal}{Astron. J.}} \textbf{\bibinfo{volume}{119}},
  \bibinfo{pages}{760--776} (\bibinfo{year}{2000}).

\bibitem{Merritt1987}
\bibinfo{author}{{Merritt}, D.}
\newblock \bibinfo{title}{{The distribution of dark matter in the coma
  cluster}}.
\newblock \emph{\bibinfo{journal}{Astrophys. J.}}
  \textbf{\bibinfo{volume}{313}}, \bibinfo{pages}{121--135}
  (\bibinfo{year}{1987}).

\bibitem{Drukier1998}
\bibinfo{author}{{Drukier}, G.~A.} \emph{et~al.}
\newblock \bibinfo{title}{{Global kinematics of the globular cluster M15}}.
\newblock \emph{\bibinfo{journal}{Astron. J.}} \textbf{\bibinfo{volume}{115}},
  \bibinfo{pages}{708--+} (\bibinfo{year}{1998}).

\bibitem{Feast1961}
\bibinfo{author}{{Feast}, M.~W.}, \bibinfo{author}{{Thackeray}, A.~D.} \&
  \bibinfo{author}{{Wesselink}, A.~J.}
\newblock \bibinfo{title}{{Analysis of radial velocities of stars and nebulae
  in the Magellanic Clouds}}.
\newblock \emph{\bibinfo{journal}{Mon. Not. R. Astron. Soc.}}
  \textbf{\bibinfo{volume}{122}}, \bibinfo{pages}{433--+}
  (\bibinfo{year}{1961}).

\bibitem{Olszewski1996}
\bibinfo{author}{{Olszewski}, E.~W.}, \bibinfo{author}{{Pryor}, C.} \&
  \bibinfo{author}{{Armandroff}, T.~E.}
\newblock \bibinfo{title}{{The Mass-to-Light Ratios of the Draco and Ursa Minor
  Dwarf Spheroidal Galaxies. II. The Binary Population and its Effects on the
  Measured Velocity Dispersions of Dwarf Spheroidals}}.
\newblock \emph{\bibinfo{journal}{Astron. J.}} \textbf{\bibinfo{volume}{111}},
  \bibinfo{pages}{750--+} (\bibinfo{year}{1996}).

\bibitem{Martin:2008wj}
\bibinfo{author}{Martin, N.~F.}, \bibinfo{author}{de~Jong, J. T.~A.} \&
  \bibinfo{author}{Rix, H.-W.}
\newblock \bibinfo{title}{{A comprehensive Maximum Likelihood analysis of the
  structural properties of faint Milky Way satellites}}
  (\bibinfo{year}{2008}).
\newblock \eprint{arXiv:0805.2945 [astro-ph]}.

\end{thebibliography}

\end{document}